\documentstyle[12pt]{article}
\newcommand{\sect}[1]{\setcounter{equation}{0}\section{#1}\indent}
\renewcommand{\theequation}{\thesection.\arabic{equation}}
\renewcommand{\thefootnote}{\fnsymbol{footnote}}
\newcommand{\EQ}{\begin{equation}}
\newcommand{\EN}{\end{equation}}
\newcommand{\bea}{\begin{eqnarray}}
\newcommand{\ena}{\end{eqnarray}}
\newcommand{\vs}[1]{\vspace{#1 mm}}

\renewcommand{\a}{\alpha}

\newcommand{\e}{\epsilon}

\newcommand{\G}{\Gamma}

\newcommand{\uda}{\nearrow \kern-1em \searrow}
\newcommand{\half}{{1 \over2}}
\newcommand{\NP}[1]{Nucl.\ Phys.\ {\bf #1}}
\newcommand{\PL}[1]{Phys.\ Lett.\ {\bf #1}}
\newcommand{\CMP}[1]{Comm.\ Math.\ Phys.\ {\bf #1}}
\newcommand{\PR}[1]{Phys.\ Rev.\ {\bf #1}}
\newcommand{\PRL}[1]{Phys.\ Rev.\ Lett.\ {\bf #1}}

\newcommand{\IJMP}[1]{Int.\ Jour.\ Mod.\ Phys.\ {\bf #1}}

\newcommand{\La}{\Lambda}

\newcommand{\mone}{$P^{1,1,2,2,6}[12]$} 
\newcommand{\mtwo}{$P^{1,1,2,2,2}[8]$}
\newcommand{\mthree}{$P^{1,1,2,2,2,2}[4,6]$}
\newcommand{\mfour}{$P^{1,1,2,2,2,2,2}[4,4,4]$}

\makeatletter
\def\eqnarray{%
 \stepcounter{equation}%
 \let\@currentlabel=\theequation
 \global\@eqnswtrue
 \global\@eqcnt\z@
 \tabskip\@centering
 \let\\=\@eqncr
 $$\halign to \displaywidth\bgroup\@eqnsel\hskip\@centering
 $\displaystyle\tabskip\z@{##}$&\global\@eqcnt\@ne
 \hfil$\displaystyle{{}##{}}$\hfil
 &\global\@eqcnt\tw@$\displaystyle\tabskip\z@{##}$\hfil
 \tabskip\@centering&\llap{##}\tabskip\z@\cr}
\makeatother

\begin{document}

\begin{titlepage}
\setcounter{page}{0}
\begin{flushright}
EPHOU 97-007\\
July 1997\\
\end{flushright}

\vs{6}
\begin{center}
{\Large  Evaluation of Periods via Fibrations in Seiberg-Witten Theories and in Type-II String}

\vs{6}
{\large
Hisao Suzuki}\\
\vs{6}
{\em Department of Physics, \\
Hokkaido
University \\  Sapporo, Hokkaido 060 Japan\\
hsuzuki@phys.sci.hokudai.ac.jp} \\
\end{center}
\vs{6}

\centerline{{\bf{Abstract}}}
 We show how to evaluate the periods in Seiberg-Witten theories and in $K3$-fibered Calabi-Yau manifolds by using fibrations of the theories. In the Seiberg-Witten theories, it is shown that the dual pair of fields can be constructed from the classical fields in a simple form.

 As for  Calabi-Yau manifolds which are fibrations of $K3$ surface, we obtain  the solutions of the Picard-Fuchs equations from the periods of $K3$ surface. By utilizing the expression of periods for two-parameter models of type-II string, we derive the solutions of the Picard-Fuchs equations around the points of enhanced gauge symmetry and  show a simple connection to the  $SU(2)$ Seiberg-Witten theory. 
\end{titlepage}
\newpage
\renewcommand{\thefootnote}{\arabic{footnote}}
\setcounter{footnote}{0}
\sect{Introduction}
Recently there has been much progress in understanding the gauge theories and string theory especially for the theories with $N=2$ supersymmetry. In the framework of the rigid theories, Seiberg-Witten theories\cite{SW} have been applied to almost all gauge groups with matter multiplets\cite{KLTY,APS}. As for the explicit evaluation of the periods, the analysis has been successfully achieved both from Picard-Fuchs equations\cite{IMNS,Alishahira} and from the explicit integration\cite{DKP,MS1,MS}.

It has been shown that  Seiberg-Witten theory\cite{SW} can be derived by the duality of the string theories\cite{KachruVafa,KKLMV,KLMVW}(For review, see refs.\cite{Lerche,Klemm}.) Recently, a systematic analysis from toric diagrams has been achieved for examination of the enhanced gauge symmetries from the conifold singularities in type-II string\cite{CF,CPR,PS,CS}.
Another interesting machinery called geometric engineering has been invented\cite{KV,KKV,KMV} starting from singular Calabi-Yau manifolds.   

As for the explicit evaluation of the periods and prepotentials for the Calabi-Yau manifolds, 
it is known that the periods can be represented by hypergeometric series\cite{Batyrev,HKTY,COFKM} in the large moduli regions.  On the other hand, as for the periods around the conifolds which are the point of phase transitions, there exists no systematic analysis except for the leading expansion which represents the Seiberg-Witten theories\cite{KKLMV,KLMVW} in $K3$-fibered Calabi-Yau manifolds.  However, it is desirable to have some systematic methods of evaluating the periods and prepotentials around these singularities for the description of the gravitationally corrected Seiberg-Witten theories. 
 For this purpose, the three-parameter models of type-II string have been studied in ref.\cite{Suzuki} where the periods are shown to be  written in the form of hypergeometric series even around the conifold singularities\cite{Suzuki}. The result serves as a first example of a systematic analysis of periods around conifold points. However, the method used there can be applied only for models which are constructed by particular sequences of fibrations. 

As for the geometrical aspects of the theories with enhanced gauge symmetries,  it is well known that Riemannian surfaces describing the geometry of $N=2$ rigid theory are constructed by the fibrations of points\cite{KLMVW,Lerche}. Correspondingly, the theories with enhanced gauge symmetries in type-II string are given by $K3$-fibrations\cite{KLM,AL}. 
These  structure suggests a systematic study  around the points of enhanced gauge symmetries both in Seiberg-Witten theory and in Type-II string via the use of the fibrations. Such usage of fibrations is expected to simplify the analysis of enhanced gauge symmetries in Type-II string.

A useful suggestion has been given in ref.\cite{LY} where they have dealt with two moduli models of Calabi-Yau manifolds. It is pointed out in ref.\cite{LY} that  half of the periods of two moduli class of Calabi-Yau manifolds  are constructed by means of the periods of one parameter family of $K3$. We can argue how to obtain the remaining half of the periods by the analyticity of the solutions. It is natural to expect that we can obtain the expression for all calabi-Yau manifolds which are $K3$-fibrations. 

Our main aim of this paper is to show a simple connection between periods of fibered manifolds and those of base manifolds. In Seiberg-Witten theories with classical gauge groups, we will obtain the dual pair of fields from the classical fields in the weak coupling region. The method may be helpful even for the study of the theories with exceptional groups. In the case of  $K3$-fibered Calabi-Yau manifolds,
 we will obtain the periods of $K3$-fibered Calabi-Yau manifolds from the periods of $K3$ surface by generalizing the expression given in ref.\cite{LY}.
 We will show that the above construction simplifies the analysis  of the periods around the point of enhanced gauge symmetries.  Namely, the expression of periods of Calabi-Yau manifolds can be obtained once  we have the expansion of periods around the singular points of $K3$ surface. This program will be performed explicitly for two-moduli models of type-II string. 

In the next section, we will show a method for obtaining dual pair of fields from the classical fields in Seiberg-Witten theory\cite{SW}.  As for the examples of the explicit evaluation, we will deal with $SU(N)$ Yang-Mills theories and find that we can construct dual pair of fields from the roots of $N$-th order equation. The method will be applicable to the theories with or without matters for classical groups. 

In section 3, we will deal with the $K3$-fibered Calabi-Yau manifolds and show that periods can be obtained from the periods of $K3$ surface. By analyzing the expansion of the periods of $K3$ around the singularities, we can obtain the expansion of the periods around the point of enhanced gauge symmetries.

In section 4, we will analyze  two-moduli models of Type-II string in detail. The models we consider are \mone, \mtwo, \mthree and \mfour. These models are known to be constructed by the fibrations over one-parameter class of $K3$.  The first model \mone {} is known to have definite dual heterotic strings\cite{KachruVafa} and the non-perturbative behavior of the model has been nicely studied\cite{KKLMV,KLMVW}. The interesting observation concerning the underlying K3 surface of the models has been made in Ref.\cite{KLM,LY}.  In these models we show how to construct the periods of the Calabi-Yau spaces out of those of underlying $K3$-surfaces both from the approach described in section 3 and from the explicit form of the periods in the large moduli regions.  By considering the expansion around the singularity of the $K3$ surface, we will construct the expansion of the periods for Calabi-Yau space.

The last section is devoted to some discussions.
\sect{Dual Pair of fields in Seiberg-Witten Theories from Classical fields}
Let us consider the relation between the vase manifolds and the fibers in the case of Seiberg-Witten theory. The curve describing the $SU(N)$ Yang-Mills theory is given by\cite{KLTY}
\bea
z+{\La^{2N}\over z} -2P(x)=0,
\ena
where $P(x)$ is defined as 
\bea
P(x) &=& x^N - s_1 x^{N-1} -s_2 x^{N-2}+\cdot\cdot\cdot -s_N\nonumber\\
     &=& \prod_{i=1}^N [x-e^i(s_k)], 
\ena
with $s_1 =0$. The period can be written as 
\bea
\omega_0^i = \oint_{\a^i} dx \oint {dz \over z} { 1\over z+{\La^{2N}\over z} -2P(x)},
\label{eq:period}\ena
where $\a^i$ represents the circle enclosing poles which shrink to $e^i(s_k)$ when $\Lambda =0$.

By expanding $(\ref{eq:period})$ with respect to $\La^{2N}$, evaluating the $z$ integral, we find that the periods can be written as
\bea
\omega_0^i = \sum_{n=0}^\infty {1 \over (n!)^2}({\La^{2N} \over 4})^n( {\partial \over \partial s_N})^{2n} \oint_{\a^i} dx {1 \over P(x)},
\ena
The last integral is exactly the periods of the classical one($\Lambda=0$ in $(\ref{eq:period})$). This expansion is valid for $\a$-cycles. Quite similarly, the integral of the meromorphic one form can be obtained by the integration of $\omega_0^i$ with respect to $s_N$. We thus find
\bea
a^i(s_k) = \sum_{n=0}^\infty {1 \over (n!)^2}({\La^{2N} \over 4})^n( {\partial \over \partial s_N})^{2n} e^i(s_k),
\label{eq:ai}\ena  
where $e^i(s_k)$ denotes a root of the equation $P(x)=0$.  

In order to obtain the expression of $a_D^i$, we will re-write $(\ref{eq:ai})$ by using Barnes-type integral representation as
\bea
a^i(s_k) = \int {ds\over 2\pi i} { \G(-s)\over \G(s+1)}(-{\La^{2N} \over 4})^s( {\partial \over \partial s_N})^{2s}e^i(s_k),
\label{eq:aibarnes}\ena
where the integral is taken to pick up the poles at integers. The formal derivative ${\partial ^{2s}\over \partial x^{2s}}F(x)$ is defined in such a way that it simply satisfies a relation:
\bea
{\partial^{2s}\over \partial x^{2s}}x^\a = {\G(\a+1)\over\G(a-2s+1)}x^{\a-2s}.
\label{eq:partialderivative}\ena
Then the expression of $a_D^i$ can be obtained by replacing ${(-1)^s/\G(s+1)}$ by $\G(-s)$:
\bea
a_D^i(s_k) ={1 \over \pi i}  \int {ds \over 2\pi i}\G(-s)^2 ({\La^{2N} \over 4})^s( {\partial \over \partial s_N})^{2s}e^i(s_k),
\label{eq:ad}\ena
where the normalization factor has been fixed by the leading behavior. Because of the logarithmic nature of the solution, the expression $(\ref{eq:ad})$ may become ambiguous when we try to use the analytic continuation. However, the ambiguity usually occurs by mod $a^i(s_k)$ so that it can be absorbed in the bare couplings.  
The validity of this expression can be shown by explicit expression of the dual pair of fields\cite{MS}. It is clear that fields $a_i$ and $a_D^i$ satisfy identical  Picard-Fuchs equations because the replacement ${(-1)^s/\G(s+1)} \rightarrow \G(-s)$ in $(\ref{eq:aibarnes})$ does not change the recursion relation satisfied by the coefficients.

Let us check the previous results in some simple examples.
 In the case of $SU(2)(P(x)=x^2-u)$\cite{SW}, the classical value $e$ is $e = {1\over \sqrt{2}}u^{1/2}$ (where we have followed the normalization of $SU(2)$), and we find
\bea
a(u) &=& {1\over \sqrt{2}}\sum_{n=0}^\infty {1\over \G(n+1)^2}({\La^2\over4})^n({d\over du})^{2n}u^{1/2},\nonumber\\
&=&{u^{1/2}\over \sqrt{2}}F(-1/4,1/4;1;\La^4/u^2).
\label{eq:su2}\ena 
It is possible to see that the expression for $a_D$ also agrees with the known expression\cite{KLT,Lerche,Klemm}. In this special case, the Seiberg-Witten curve can be embedded in $P^{1,1,2}[4]$ which is the fiber of the points $P^{1,1}[2]$.

For the group $SU(3)$, the root of the third order equation $x^3 -ux -v=0$ can be solved algebraically. In the region ${v^2\over u^3} << 1$, one can express the roots by using hypergeometric functions as
\bea
e_1 &=& -{v\over u}F({1\over3},{2\over3};{3\over2};{27\over4}{v^2\over u^3}),\nonumber\\
e_2 &=& \sqrt{u}F(-{1\over6},{1\over6};{1\over2};{27\over4}{v^2\over u^3})+{v\over 2u}F({1\over3},{2\over3};{3\over2};{27\over4}{v^2\over u^3}),\nonumber\\
e_3 &=& -\sqrt{u}F(-{1\over6},{1\over6};{1\over2};{27\over4}{v^2\over u^3})+{v\over 2u}F({1\over3},{2\over3};{3\over2};{27\over4}{v^2\over u^3}).
\label{eq:su3}\ena
It is easy to see that dual pair of fields constructed by the above formula $(\ref{eq:ai})$ and $(\ref{eq:ad})$ can be written in terms of Appell functions\cite{Appell}, which agrees with the known results\cite{KLT,MS1,MS,MSS}. For the $SU(N)$ case, the explicit forms of the dual pair of fields are listed in ref.\cite{MS}. 

It is straightforward to generalize these expressions to other classical gauge groups with or without matters by suitable inclusions of source terms, which clearly shows a characteristic property of the simple base spaces and the fibers.  In addition, we can obtain the similar expression even for ALE-fibrations\cite{KLTY,LW} although the  correspondence to the classical point set is not manifest\cite{LW}. We hope that the  above construction will be helpful for the analysis of the theories with these exceptional groups as well as the studies of the theories obtained in ref.\cite{Witten}.
\sect{Periods through fibrations for $K3$-fibered Calabi-Yau Manifolds}
In the previous section, we have shown how to obtain the dual pair of fields from the classical fields in Seiberg-Witten theories.
We can expect the same kind of expression of periods for $K3$ fibered Calabi-Yau manifolds. Namely, the periods of Calabi-Yau manifolds can be expressed by those of the $K3$ surface.  A simple expression has been obtained for two-parameter models\cite{LY}.  We are going to generalize the expression used for two-moduli models\cite{LY}.

Let us consider the models of type  $P^{1,1,2k_1,2k_2,2k_3}[2d],(d=1+k_1+k_2+k_3)$ which are fibrations of the $K3$ surface $P^{1,k_1,k_2,k_3}[d]$. The defining polynomial can be written in the form:
\bea
f_{CY} = {y_0 \over 2d}[z_1^{2d} +z_2^{2d} + {2\over\sqrt{y_s}}(z_1z_2)^d] + \hat{W}({z_1z_2\over y_s^{1/2d}},z_k;y_i),
\ena
where $y_0,y_s$ and $y_i$ denote moduli of the manifold. We can always find this form by the suitable redefinition of the moduli variables from the one listed in ref.\cite{KLMVW,Lerche}.

By a change of variables to $z_1 = \sqrt{z_0}\zeta^{1\over 2d}, z_2 = \sqrt{z_0}\zeta^{-{1\over 2d}}y_s^{1\over 2d}$, we rewrite the defining polynomial as follows:
\bea
f_{CY} = {y_0\over 2d}(\zeta+{y_s\over\zeta})z_0^d + f_{K3},
\label{eq:defcy}\ena
where $f_{K3}$ is the defining polynomial of the base $K3$ surface:
\bea
f_{K3} = {y_0 \over d}z_0^d + \hat{W}(z_0,z_k:y_i).
\label{eq:defk3}\ena
Note that the form $(\ref{eq:defcy})$ is almost identical to the curve $(2.1)$ in Seiberg-Witten theory. It is now straightforward to repeat the calculation for the periods of Calabi-Yau manifold as follows:
\bea
\omega &=& \oint{dz\over z} \int {d\Omega \over f_{CY}},\nonumber\\
&=& \sum_{n=0}^\infty{1 \over (n!)^2} ({y_sy_0^2 \over4})^n ({d\over dy_0})^{2n} \int {d\Omega \over f_{K3}},
\label{eq:calabi-yau}\ena
where $d\Omega$ represents the measure of $K3$ surface.

Therefore we find that that the half of the periods can be obtained by means of $K3$ periods. The remaining half of the solutions can be obtained by analyticity of the form of solutions as has been given in the previous section. Namely, we can formally write the solutions of the Picard-Fuchs equations of Calabi-Yau $(\ref{eq:calabi-yau})$ in the form:
\bea
\int {ds \over 2\pi i}{\G(-s)\over\G(s+1)}(-{y_sy_0^2 \over4})^s({\partial \over \partial y_0})^{2s} F(y_0,y_i),
\label{eq:solgeneral}\ena
where $F(y_0,y_i)$ is the solutions of the Picard-Fuchs equations of $K3$ surface whose defining equation is given by $f_{K3}=0$. Then the remaining half of the solutions can be written by
\bea
\int {ds \over 2\pi i}\G(-s)^2({y_sy_0^2 \over4})^s({\partial \over \partial y_0})^{2s} F(y_0,y_i).
\label{eq:solgenerallog}\ena

In this way, we can derive the solutions of the Picard-Fuchs equations for $K3$-fibered Calabi-Yau manifold out of the solutions for $K3$ surface. It is also possible to generalize the above construction for more generic type of models of  $P^{1,l,(l+1)k_1,(l+1)k_2,(l+1)k_3}[(l+1)d].$
\sect{Construction of periods in two-moduli models of type-II string around the point of enhanced gauge symmetry}
Let us consider the Type-II string compactified on the Calabi-Yau manifold with $h_{11}=2$. A typical example for the enhanced gauge symmetry is the $K3$-fibered threefold $P^{1,1,2,2,6}[12]$. In the type-IIB side, the defining polynomial is given by\cite{HKTY,COFKM}
\bea
f= {1\over12} z_1^{12} + {1 \over 12}z_2^{12} + {1\over6}z_3^{6}&+&{1\over6}z_4^{6}+{1\over2}z_5^{2} \nonumber\\
&-&\psi_0z_1z_2z_3z_4z_5 +{1\over6}\psi_1(z_1z_2)^6,
\label{eq:eq21}\ena
where we have followed by the normalization given in Ref.\cite{KKLMV,KLMVW}.

This model is known to be the fiber of a $K3$ surface $P^{1,1,1,3}[6]$, which represents one-parameter class of the $K3$ surface. In order to cast the defining equation into the form $(\ref{eq:defcy})$, we re-scale the variables $z_1$ and $z_2$ to find 
\bea
f= {x \over 12}[z_1^{12}+z_2^{12} +{2 \over \sqrt{y}}(z_1z_2)^6]+{1\over6}z_3+{1\over6}z_4^6+\half z_5^2 - (y^{-1/12}z_1z_2)z_3z_4z_5,
\ena
where we have defined $x=\psi_1/\psi_0^6$ and $y=1/\psi_1^2$. Therefore, we can construct the solutions of the Picard-Fuchs equations of the model out of the solutions for $K3$ surface whose defining polynomial is $f_{K3} = {x \over 6} z_0^6 + {1\over6}z_3^6 +{1 \over6}z_4^6 + \half z_5^2 - z_0z_3z_4z_5$ which is equivalent to the standard form $f_{K3} = {1\over 6} z_0^6 + {1\over6}z_3^6 +{1 \over6}z_4^6 - {\tilde{\psi}}z_0z_3z_4z_5,(x={1/{\tilde{\psi}^6}}).$ 
It is known that there are three other models having this kind of property\cite{KLM,LY}. That is, the models are \mtwo, \mthree and \mfour, which are fiber over $P^{1,1,1,1}[4], P^{1,1,1,1,1}[2,1]{}$ and $P^{1,1,1,1,1,1}[2,2,2]$, respectively. The defining polynomials in the type-IIB side are given by\\
\noindent\mtwo: 
\bea
f={1\over 8} z_1^8 + {1\over8}z_2^8 + {1\over4}z_3^4 &+& {1\over4}z_4^{4}+ {1\over4}z_5^{4}\nonumber\\
 &-& \psi_0z_1z_2z_3z_4z_5 + {1\over4}\psi_1(z_1z_2)^4,
\label{eq:eq22}\ena
\noindent\mthree: 
\bea
f_1&=&{1\over4} z_1^4 + {1\over4}z_2^4 + {1\over2}z_3^2 + {1\over2}z_4^{2}
 - z_5z_6 + {1\over2}\psi_1(z_1z_2)^2,\nonumber\\
f_2 &=& {1\over3}z_5^3 +{1\over3}z_6^3 - \psi_0z_1z_2z_3z_4,
\label{eq:eq23}\ena
\noindent\mfour:
\bea
f_1 &=& {1\over4}z_1^4 + {1\over4}z_2^4+{1\over2}z_3^2 - z_4z_5 +\half \psi_1(z_1z_2)^2, \nonumber\\
f_2 &=& \half z_4^2 + \half z_5^2 - z_6z_7,\nonumber\\
f_3 &=& \half z_6^2  +\half z_7^2 - \psi_0 z_1z_2z_3.
\label{eq:eq24}\ena

Following the notation of Ref.\cite{KKLMV}, we introducing the parameters by $x=\psi_1/\psi_0^{1\over\lambda}, y={1/\psi_1^2}$, where $\lambda$ is $1/6, 1/4, 1/3$ and $1/2$ for \mone, \mtwo, \mthree{} and \mfour, respectively. From the general construction given in the previous section, we can construct the periods of these Calabi-Yau manifolds from those of $K3$ surface. We are going to check the formula in the previous section from the solutions of Picard-Fuchs equations in the large moduli limit where the periods of the models are known well\cite{COFKM,HKTY}.
At large moduli $x<1,y<1$, the fundamental period of the system can be obtained as
\bea
\omega_0 =
\sum_{n,m =0}^\infty
 {\G(n+\lambda)\G(n+1-\lambda)\G(n+\half)  \over \G(n+1)^2\G(m+1)^2\G(n-2m+1)} x^n({y\over 4})^m,
\ena
which can be re-written 
 in the form:
\bea
\omega_0= \int{ds_1\over 2\pi i}\int{ds_2\over 2\pi i}{\G(-s_1)\G(-s_2)\G(s_1+\lambda)\G(s_1+1-\lambda)\G(s_1+\half) \over \G(s_1+1)\G(s_2+1)\G(s_1-2s_2+1)}(-x)^{s_1}(-{y\over 4})^{s_2},
\label{eq:eq27}\ena
where the contour integrals are  taken to be circles enclosing integers. The other solutions can be obtained by the replacements such as $1/\G(s_1+1) \rightarrow (-1)^{s_1}\G(-s_1),1/\G(s_2+1) \rightarrow (-1)^{s_2}\G(-s_2) $ in $(\ref{eq:eq27})$.  In order to clarify the role of the $K3$-fibrations, we will cast these solutions in the form (\ref{eq:solgeneral}) and (\ref{eq:solgenerallog}):
\bea
&{}&\sum_{m=0}^\infty{ 1\over \G(m+1)^2}({x^2y\over 4})^m{d^{2m}\over dx^{2m}}F(x),\label{eq:soloftwoone}\\
&{}&\int{ds\over 2\pi i}\G(-s)^2({x^2y\over 4})^s{d^{2s}\over dx^{2s}}F(x),
\label{eq:soloftwo}\ena
where $F(x)$ is the periods of the $K3$-surface which are solutions of the Picard-Fuchs equation:
\bea
[\theta_x^3 -x(\theta_x+\half)(\theta_x+\lambda)(\theta_x+1-\lambda)]F(x)=0,
\label{eq:pfk3}\ena
where $\theta_x \equiv x d/dx$.

In $(\ref{eq:soloftwo})$, the formal derivative ${d^{2s}\over dx^{2s}}F(x)$ is defined in such a way that it simply satisfies the relation $(\ref{eq:partialderivative})$.
Since the equation $(\ref{eq:pfk3})$ have three independent solutions, the solutions $(\ref{eq:soloftwo})$ represent six independent solutions of the Picard-Fuchs equations of the $K3$-fibered Calabi-Yau manifolds. The expression of the periods $(\ref{eq:soloftwoone})$ can be found in ref.\cite{LY}. In the previous section,  we have shown that the  similar expansion can be extended for all Calabi-Yau manifolds with $K3$-fibrations as well as the construction of the logarithmic solutions $(\ref{eq:soloftwo})$. 

Let us analyze the solutions of $(\ref{eq:pfk3})$ around the singularity $x=1$. An interesting observation has been made for these one-parameter class of $K3$ surfaces\cite{LY}. Namely, let $f(x)$ and $g(x)$ be the solutions of a hypergeometric equation:
\bea
[\theta_x^2-x(\theta_x+{\lambda\over2})(\theta_x+{1-\lambda\over2})]f=0,
\label{eq:second}\ena
then, the equation $(\ref{eq:pfk3})$ are solved by $F(x) = f(x)g(x)$. 
We can prove this result by considering
\bea
\theta_x^3(fg) = [g\theta_x+3(\theta_xg)]\theta_x^2f + [f\theta_x+3(\theta_xf)]\theta_x^2g,
\label{eq:inter}\ena
By using $(\ref{eq:second})$, we can find that right hand side of $(\ref{eq:inter})$ can be written as 
$x(\theta_x+\half)(\theta_x+\lambda)(\theta_x+1-\lambda)(fg)$, which proves the result\cite{LY}. This statement itself is a very old result which can be found in the standard textbook\cite{WW}.  
A classical identity ${}_3F_2(\lambda, 1-\lambda,\half;1,1:x) = [{}_2F_1(\lambda/2,(1-\lambda)/2;1;x)]^2 $ is a direct consequence of the result\cite{WW}. We can find identities for the logarithmic solutions. What we need here is the expansion around singularities.  The two independent solutions of the equation $(\ref{eq:second})$ around $x=1$ are given by ${}_2F_1(\lambda/2,(1-\lambda)/2;1/2;1-x)$ and $(1-x)^\half{}_2F_1(1-\lambda/2,1/2-\lambda/2;3/2;1-x)$. The basic solutions of the Picard-Fuchs equation of $K3$ are obtained by products of these functions. It is not difficult to obtain the periods by the analytic continuation from those in the large moduli $(x=0)$, which will not be pursued here. We are going to  analyze the behavior of the solutions around the conifold locus directly.  

The basic solutions of the Picard-Fuchs equations of the Calabi-Yau manifolds can be obtained from $(\ref{eq:soloftwo})$ in the form:
\bea
&{}&\sum_{m,n_1,n_2=0}^\infty{\G(n_1+{\lambda\over2})\G(n_1+{1-\lambda\over2})\G(n_2+{\lambda\over2})\G(n_2+{1-\lambda\over2})\G(n_1+n_2+1)\over\G(n_1+\half)\G(n_1+1)\G(n_2+\half)\G(n_2+1)\G(n_1+n_2-2m+1)\G(m+1)^2}\nonumber\\
&{}&\qquad\qquad\qquad \times ({x^2y\over4(x-1)^2})^m(1-x)^{n_1+n_2},\label{eq:asol}\\
&{}&\sum_{m,n_1,n_2=0}^\infty{\G(n_1+{\lambda\over2})\G(n_1+{1-\lambda\over2})\G(n_2+1-{\lambda\over2})\G(n_2+{1+\lambda\over2})\G(2m-n_1-n_2-\half)\over\G(n_1+\half)\G(n_1+1)\G(n_2+{3\over2})\G(n_2+1)\G(-n_1-n_2-\half)\G(m+1)^2}\nonumber\\
&{}&\qquad\qquad\qquad\times({x^2y\over4(x-1)^2})^m(x-1)^{n_1+n_2+\half},\label{eq:bsol}\\
&{}&\sum_{m,n_1,n_2=0}^\infty{\G(n_1+1-{\lambda\over2})\G(n_1+{1+\lambda\over2})\G(n_2+1-{\lambda\over2})\G(n_2+{1+\lambda\over2})\G(n_1+n_2+2)\over\G(n_1+{3\over2})\G(n_1+1)\G(n_2+{3\over2})\G(n_2+1)\G(n_1+n_2-2m+2)\G(m+1)^2}\nonumber\\
&{}&\qquad\qquad\qquad({x^2y\over4(x-1)^2})^m(1-x)^{n_1+n_2+1},\label{eq:csol}
\ena
The logarithmic solutions can be obtained by use of the second form of $(\ref{eq:soloftwo})$. It seems difficult to obtain these compact expressions of solutions directly from the Picard-Fuchs equations of Calabi-Yau manifolds\cite{COFKM,HKTY,KKLMV}.

When we use the variables\bea
x=1-\e u, \qquad y={\e^2 \Lambda^4 },
\ena
we find that the leading order of the solution $(\ref{eq:bsol})$ can be written as
\bea
u^{\half}F(-{1\over4},{1\over4};1;{\Lambda^4\over u^2}),
\ena
which agrees with $(\ref{eq:su2})$; the solution of the Picard-Fuchs equation of the $SU(2)$ Seiberg-Witten theory. The leading behavior of the solution for the Calabi-Yau manifolds can be easily obtained from the leading behavior of a solution of the Picard-Fuchs equation for $K3$ surface: $F(x) \sim (1-x)^{1/2} \sim u^{1/2}.$ It is interesting that this result of enhanced gauge symmetries observed in ref.\cite{KKLMV} for \mone{} can be observed for the solution space of the Picard-Fuchs equation in a very simple form.  

Strictly speaking, the expansion of the solutions $(\ref{eq:asol}),(\ref{eq:bsol})$ and $(\ref{eq:csol})$ are not in the hypergeometric form with respect to the variable $1-x$. The coefficients can be expressed by well-poised ${}_4F_3$\cite{HTF}. This is in contrast to the case of the three-parameter models which can be written in the form of hypergeometric series\cite{Suzuki}. The difference can be explained by the fact that the one-moduli models of $K3$ surface are regarded as one-parameter family of two-parameter models of $K3$ surfaceof ref.\cite{LY,Suzuki}, which can be found at the level of the explicit solutions, although we could not prove the statement directly from the defining polynomials.  

\sect{Discussions}
We have shown that the evaluation of periods can be simplified when we use the structure of fibrations both in Seiberg-Witten theories and in $K3$-fibered Calabi-Yau manifolds. For Seiberg-Witten theories, the method may be useful for the analysis of exceptional groups or the theories considered in ref.\cite{Witten,KMV}. As for the analysis of the Type-II strings, we have shown that periods of Calabi-Yau manifolds can be obtained once we have the periods of $K3$ surface. However, at present, the simple expression around the singularities for such $K3$ are available only for one and two moduli class of models. It is known that the classification of the leading singularities for $K3$-surface can be performed. However, it seems important to analyze expansion of periods around the singular points for generic $K3$ surface for the study of the enhanced gauge symmetries with the gravitational corrections.


\end{document}